\documentclass[letterpaper,twocolumn,english,showpacs,preprintnumbers,amsmath,amssymb]{revtex4-1}
\usepackage[latin9]{inputenc}
\usepackage{amsmath}
\usepackage{amssymb}
\usepackage{graphicx}
\usepackage{esint}

\makeatletter

 \@ifundefined{textcolor}{}
 {%
   \definecolor{BLACK}{gray}{0}
   \definecolor{WHITE}{gray}{1}
   \definecolor{RED}{rgb}{1,0,0}
   \definecolor{GREEN}{rgb}{0,1,0}
   \definecolor{BLUE}{rgb}{0,0,1}
   \definecolor{CYAN}{cmyk}{1,0,0,0}
   \definecolor{MAGENTA}{cmyk}{0,1,0,0}
   \definecolor{YELLOW}{cmyk}{0,0,1,0}
 }

\usepackage{epsfig}\usepackage{subfigure}\usepackage{dcolumn}\usepackage{stmaryrd}\usepackage{mathrsfs}\usepackage{pifont}\usepackage{amsthm}\usepackage{bm}\usepackage{latexsym}\usepackage{amsfonts}

\usepackage{babel}

\usepackage{babel}

\makeatother

\usepackage{babel}
\begin{document}

\title{Phonon mediated spin relaxation in a moving quantum dot: Doppler shift, Cherenkov radiation, and spin relaxation boom}

\author{Xinyu Zhao$^{1}$}

\email{xzhao34@buffalo.edu}

\author{Peihao Huang$^{1,2}$}

\email{peihao.huang@csun.edu}

\author{Xuedong Hu$^{1}$}

\email{xhu@buffalo.edu}

\affiliation{$^{1}$Department of Physics, University at Buffalo, SUNY, Buffalo,
New York 14260, USA}

\affiliation{$^{2}$Department of Physics, California State University Northridge,
Northridge, California 91330, USA}
\begin{abstract}
We study relaxation of a moving spin qubit caused by phonon noise.
As we vary the speed of the qubit, we observe several interesting
features in spin relaxation and the associated phonon emission, induced
by Doppler effect. In particular, in the supersonic regime, the phonons
emitted by the relaxing qubit is concentrated along certain directions,
similar to the shock waves produced in classical Cherenkov effect.
As the speed of the moving qubit increases from the subsonic regime
to the supersonic regime, the qubit experiences a peak in the spin
relaxation rate near the speed of sound, which we term a spin relaxation
boom in analogy to the classical sonic boom. We also find that the
moving spin qubit may have a lower relaxation rate than a static qubit,
which hints at the possibility of coherence-preserving transportation
for a spin qubit. While the physics we have studied here has strong
classical analogies, we do find that quantum confinement for the spin
qubit plays an important role in all the phenomena we observe. Specifically,
it produces a correction on the Cherenkov angle, and removes the divergence
in relaxation rate at the sonic barrier. It is our hope that our results
would encourage further research into approaches for transferring
and preserving quantum information in spin qubit architectures.
\end{abstract}


\maketitle

\section{Introduction}

Electron spin qubit is a promising candidate for realizing quantum
computing because of its long coherence time \cite{Amasha_PRL08,Balasubramanian_NMat09,Bluhm_NPhys11,Muhonen_NNano14}.
It has attracted extensive research interests over the past decade,
with studies mostly focusing on the fabrication and manipulation of
spin qubits confined in a fixed quantum dot or dopant ion \cite{Hanson_RMP07,Morton_Nature11,Awschalom_Science13}.

In a large-scale quantum information processor, it is inevitable that
quantum information is transferred over finite distances frequently.
One straightforward way to achieve such communication is to move the
qubits themselves directly. There are several proposed schemes on
how to move spin qubits efficiently \cite{SAW6, Gate1,Gate3,Gate2,SAW1,SAW2,SAW3,SAW4,SAW5},
where the motion of the confined electron can be induced by either
varying gate voltages or a surface acoustic wave (SAW). However, introducing
this orbital (albeit controlled) dynamics could weaken the orbital
quantization that gives rise to the long spin coherence times. For
instance, in Ref.~\cite{Huang_PRB_2013} we have shown how electrostatic
disorder in the substrate may cause relaxation of a moving spin qubit
through spin-orbit interaction. Nevertheless, more studies are still
needed to clarify decoherence of a moving spin qubit.

\begin{figure}
\noindent \begin{centering}
\includegraphics[width=1\columnwidth]{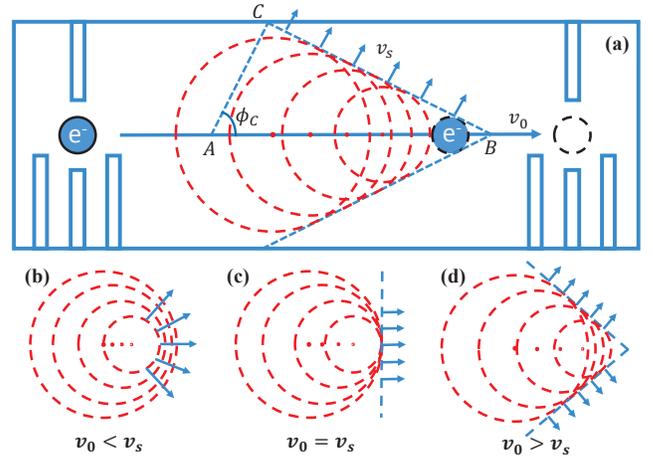}
\par\end{centering}

\caption{\label{fig:Setup}(color online) A schematic diagram of a moving spin
qubit interacting with phonon reservoir and the resultant Doppler
effect in the three cases.}
\end{figure}

Doppler effect is a commonly observed phenomenon when an object is
moving, where an observer hears different pitches from the horn of
an approaching and a departing vehicle. When the velocity of the object
is larger than the speed of the waves produced by the motion, a directional
shock wave (Cherenkov effect) can be observed, from the wake of a
speeding boat, to the sonic boom from a supersonic airplane, to Cherenkov
radiation from a fast-moving charge in a material with high refraction
index \cite{CR1,CR2,CR3,CR4}. Classically (as depicted in Fig.~\ref{fig:Setup}),
a superposition of the spherical waves emitted by a moving object
at different moments form a straight-line wavefront $BC$, and the
wavefront propagates in the $AC$ direction, at the Cherenkov angle
from the motion direction:
\begin{equation}
\cos\phi_{C}=\frac{AC}{AB}=\frac{v_{s}}{v_{0}}.\label{eq:AngleCh}
\end{equation}
Since Cherenkov radiation is highly directional, it is often used
to detect properties of moving charged particles.

Here we study how motion of a spin qubit could modify its relaxation
due to spin-orbit interaction and electron-phonon interaction \cite{SO1,Golovach_PRL_2004,Phonon2,Phonon3}.
In particular, we identify different regimes of quantum dot moving
velocity where we can find analogues of Doppler effect, Cherenkov
radiation, and sonic boom in the spin relaxation and the associated
phonon emission. More specifically, when the quantum dot (QD) moves
with a speed lower than the speed of sound, the energy of an emitted
phonon is dependent on the direction of emission, similar to the Doppler
effect. When the QD moves faster than the speed of sound, the dominant
contributions to spin relaxation come from phonons emitted along certain
directions, similar to the classical Cherenkov effect. Our calculation
predicts a small correction to the Cherenkov angle caused by the quantum
confinement. In the transition from subsonic to supersonic regime,
we observe a peak in spin relaxation rate, which we term as a spin
relaxation boom in analogy to the classical sonic boom.

The rest of the paper is organized as follows. In Section II we present
our theoretical model and the derived spin relaxation rate. In Section
III we analyze the angular distribution of the emitted phonons, focusing
on the Cherenkov effect of directional phonon emission and effects
of quantum confinement. In Section IV we clarify the overall spin
relaxation in the different regimes of QD motion, with particular
focus on spin relaxation boom and motion-dependence of the spin relaxation
rate. In Section V we discuss the implications of our results, and
in Section VI we present our conclusions. In addition, in the Appendices
we give brief summaries of our theoretical derivations with regard
to spin-orbit interaction in the context of a moving spin qubit.

\section{Model and solution}

The system we consider is a single electron confined in a moving QD
formed from a two-dimensional electron gas (2DEG), as is shown in
Fig.~\ref{fig:Setup}. The qubit (electron) is moved at a constant
speed $v_{0}$, presumably achieved by programming the gates or using
the surface acoustic waves. Conceptually, to ensure such a uniform
linear motion for the electron, there has to be an external driving
force, which we treat as a classical force. The total Hamiltonian
\cite{Golovach_PRL_2004,Phonon3,Huang_PRB_2013} is given by
\begin{equation}
H=H_{d}+H_{Z}+H_{SO}+U_{ph}(r).\label{eq:Htot}
\end{equation}
Here $H_{d}=\frac{\pi^{2}}{2m^{*}}+U[r-r_{0}(t)]$ is the orbital
Hamiltonian for the moving QD, where $\pi=-i\hbar\nabla+(e/c)A(r)$
is the 2D momentum operator of the electron, and $m^{*}$ is the effective
mass of the electron. The motion we considered is linear: $r_{0}(t)=v_{0}t$,
and the QD confinement potential $U(r-r_{0})=\frac{1}{2}m^{*}\omega_{d}^{2}(r-r_{0})^{2}$
is quadratic. $H_{Z}=\frac{1}{2}g\mu_{B}B_{0}\cdot\sigma$ is the
Zeeman Hamiltonian, with $B_{0}$ the applied magnetic field. $H_{SO}=\beta_{-}\pi_{y}\sigma_{x}+\beta_{+}\pi_{x}\sigma_{y}$
is the spin-orbit (SO) interaction, where $\beta_{\pm}=\beta\pm\alpha$
give the SO coupling strength, with $\alpha$ and $\beta$ being the
strengths of Rashba \cite{Bychkov_JPC_1984} and Dresselhaous \cite{Dresselhaus_PR_1955}
SO interaction, respectively. Lastly, the electron-phonon interaction
is given by \cite{Golovach_PRL_2004,Phonon3,Mahan}
\begin{equation}
U_{ph}(r)=\sum_{qj}\frac{F(q_{z})e^{iq_{\parallel}\cdot r}}{\sqrt{2\rho_{c}\omega_{qj}/\hbar}}\left(e\beta_{qj}-iq\Xi_{qj}\right)\left(b_{-qj}^{\text{\ensuremath{\dagger}}}+b_{qj}\right),\label{eq:Uph}
\end{equation}
where $b_{qj}^{\dagger}$ and $b_{qj}$ are the creation and annihilation
operators for an acoustic phonon with wave vector $q=(q_{\parallel},q_{z})$
and branch index $j$, and $\rho_{c}$ is the density of the material.
The function $F(q_{z})=\exp\left(-\frac{q_{z}^{2}}{2d^{2}}\right)$
models the confinement along $z$ direction, where $d$ is the characteristic
width of the quantum well. We take into account both piezoelectric
potential ($\beta_{qj}$) and deformation potential ($\Xi_{qj}$)
in the electron-phonon interaction \cite{DFPZ,Mahan}. By performing
a Schrieffer-Wolff transformation to remove the SO coupling term to
the first order \cite{Golovach_PRL_2004,Phonon2,Phonon3,Huang_PRB_2013,Peihao14},
which we briefly summarize in Appendix \ref{sec:Heff}, the effective
spin Hamiltonian can be obtained
\begin{equation}
H_{eff}=\frac{1}{2}g\mu_{B}[B_{0}+\Delta B+\delta B(t)]\cdot\sigma,\label{eq:Heff}
\end{equation}
where $\Delta B=\frac{2m^{*}}{g\mu_{B}}(\beta_{-}v_{0y},\beta_{+}v_{0x},0)$
is a motion induced constant magnetic field for the spin, and $\delta B(t)=2B_{0}\times\Omega(t)$
is the motion induced magnetic noise, where the time-dependent function
$\Omega(t)$ originates from the phonon environment
\begin{equation}
\Omega(t)=\left\langle \psi\left|\frac{-1}{\hbar\omega_{d}^{2}}\left[\beta_{-}\frac{\partial U_{ph}}{\partial y},\ \beta_{+}\frac{\partial U_{ph}}{\partial x},\ 0\right]\right|\psi\right\rangle .
\end{equation}
Here $|\psi\rangle$ is the instantaneous orbital ground state of
the QD, so that $\langle\psi|\exp(iq\cdot r)|\psi\rangle=\exp\left[iq\cdot r_{0}(t)\right]e^{-q^{2}\lambda^{2}/4}$,
where $\lambda^{-2}=\hbar^{-1}\sqrt{(m^{*}\omega_{d}^{2})^{2}+(eB_{z}/2c)^{2}}$
is the total confinement length of the QD.

With the effective Hamiltonian (\ref{eq:Heff}), the spin relaxation
rate can be obtained as (the detailed derivation is summarized in
Appendix \ref{sec:DecoRate})
\begin{equation}
\frac{1}{T_{1}}=\int_{0}^{\pi}d\theta\int_{0}^{2\pi}d\phi f(\theta,\phi),\label{eq:T1}
\end{equation}
where
\begin{equation}
f=\sum_{j}\frac{\hbar\omega_{Z}F_{SO}}{(m^{*}\omega_{d}^{2})^{2}}\frac{\left(2N_{w_{z}}+1\right)}{8\pi^{2}\rho_{c}v_{j}^{4}}w_{z}^{4}\sin^{3}\theta\cos^{2}\phi C_{ep}F_{z}F_{xy},\label{eq:fj}
\end{equation}
where
\begin{equation}
F_{z}=\exp\left(-\frac{d^{2}w_{z}^{2}}{v_{j}^{2}}\cos^{2}\theta\right),\; F_{xy}=\exp\left(-\frac{\lambda^{2}w_{z}^{2}}{2v_{j}^{2}}\sin^{2}\theta\right),\label{eq:Fc}
\end{equation}
are the cutoff functions in $z$ direction and $xy$ plane, respectively.
They reflect the quantum confinement effect that will be discussed
in the next section. The constant $C_{ep}=\left(e^{2}\beta_{qj}^{2}+\frac{w_{z}^{2}}{v_{j}^{2}}\Xi_{qj}^{2}\right)$
gives the total strength of the two types of electron-phonon interaction,
namely the deformation potential and the piezoelectric potential.
$N_{w_{z}}=\left(e^{\hbar w_{z}/T}-1\right)^{-1}$ is the number of
phonons with frequency $w_{z}$ at thermal equilibrium. The factor
$F_{SO}$ in Eq.~(\ref{eq:fj}) describes the angular dependence
of the magnetic noise on the direction of the applied field, which
can be expressed as $F_{SO}=(\beta^{2}+\alpha^{2})(1+\cos^{2}\theta_{B})+2\alpha\beta\sin^{2}\theta_{B}\cos(2\varphi_{B})$.
Lastly, the angular dependence of kernel function $f$, and therefore
the spin relaxation rate $1/T_{1}$, depends on a direction-dependent
``shifted frequency'' for the phonons,
\begin{equation}
w_{z}=\left|\frac{\omega_{Z}}{1-\xi_{j}}\right|\,,\label{eq:modfreq}
\end{equation}
instead of the spin Zeeman splitting $\omega_z=g\mu_BB_0/\hbar$. Here $\xi_{j}=\frac{v_{0}}{v_{j}}\sin\theta\cos(\phi-\phi_{v})$.
This is the Doppler shift in the context of moving spin relaxation.

In this model, spin relaxation is caused by the interaction between
the electron and phonons from all directions. The double integration
over $\theta$ and $\phi$ in Eq.~(\ref{eq:T1}) originates from
the summation $\sum_{qj}$ over all the phonon wave vectors $q$ in
Eq.~(\ref{eq:Uph}). Therefore, the kernel function $f(w_{z},\theta,\phi)$
describes contributions by phonons emitted or absorbed in the infinitesimal
solid angle $d\theta d\phi$ around $(\theta,\phi)$. In our numerical
calculations, we use typical parameters in a GaAs QD. There is one
branch of longitudinal acoustic (LA) phonons, and two branches of
transverse acoustic (TA) phonons. $v_{1}=4730$ m/s is the sound speed
of the LA phonons, while $v_{2}=v_{3}=3350$ m/s are the sound speed
of the TA phonons. The strength of the deformation potential is $\Xi_{1}=6.7$ eV.
The strengths of the piezoelectric interaction are $\beta_{1}(\theta)=3\sqrt{2}\pi h_{14}\kappa^{-1}\sin^{2}\theta\cos\theta$,
$\beta_{2}(\theta)=\sqrt{2}\pi h_{14}\kappa^{-1}\sin2\theta$, and
$\beta_{3}(\theta)=\sqrt{2}\pi h_{14}\kappa^{-1}(3\cos^{2}\theta-1)$,
where $h_{14}=-0.16$ C/m$^{2}$ and $\kappa=13.1$ \cite{Golovach_PRL_2004}.

With the help of the analytical expression of the relaxation rate,
in the next two Sections we examine in detail the features of the
angular dependence of the kernel function $f$ and the total relaxation
rate $1/T_{1}$ for the moving spin qubit.

\section{Directional Phonon Emission: Doppler effect and Cherenkov Radiation}

In this Section we analyze the angular dependence of the phonon emission
(in terms of the kernel function $f$) from the relaxing spin qubit
in different regimes of QD moving speed. In particular, in the subsonic
regime, we find the Doppler effect, in which phonons emitted in different
directions have different frequencies. In the transonic regime we
find the formation of a shock wave front and its bifurcation into
two directions as the QD speed passes the speed of sound. Lastly in
the supersonic regime we find a phonon analog of Cherenkov radiation,
and identify a quantum confinement induced correction in the Cherenkov
angle.

\subsection{\label{sub:Doppler}Doppler effect}

When a QD moves relative to the lattice with a speed smaller than
the speed of sound, the frequency of the phonon emitted or absorbed
is shifted with a Doppler factor $\frac{1}{1-\xi_{j}}$, as indicated
in Eq.~(\ref{eq:modfreq}). In particular, in the forward direction
($\phi-\phi_{v}=0$ and $\theta=\pi/2$), an emitted phonon has an
increased frequency $\omega_{Z}/(1-v_{0}/v_{j})$, while in the backward
direction the phonon frequency is reduced to $\omega_{Z}/(1+v_{0}/v_{j})$.
These shifts are exactly as one would find in the classical Doppler
effect.

It may seem puzzling that the energy quantum carried by the emitted
phonon is not the same as the Zeeman splitting of the spin qubit.
The discrepancy here can be accounted for by the fact that the moving
quantum dot is an open system. It is driven by a classical force that
comes from either programmed gate potential or the large number of
phonons in an SAW. The excess or shortage of energy in the spin relaxation
is absorbed/added by the classical ``reservoir''.

\subsection{\label{sub:boom}Breaking the sound barrier}

If the moving spin qubit acts classically, the transition from subsonic
regime to supersonic regime (the transonic regime) for the moving
spin qubit would be well represented by Fig.~\ref{fig:Setup} (b),
(c), and (d). At low speeds, presented in panel (b), there is no strongly
directional emission. As the QD moving velocity becomes equal to the
sound velocity, as indicated in panel (c), a single forward-propagating
shock wave front is formed. When the moving velocity is larger than
the critical velocity (d), the single shock wave front splits into
two (We only consider the $x-y$ plane. In 3D the wave front
is conical).

\begin{figure}
\noindent \includegraphics[width=1\columnwidth]{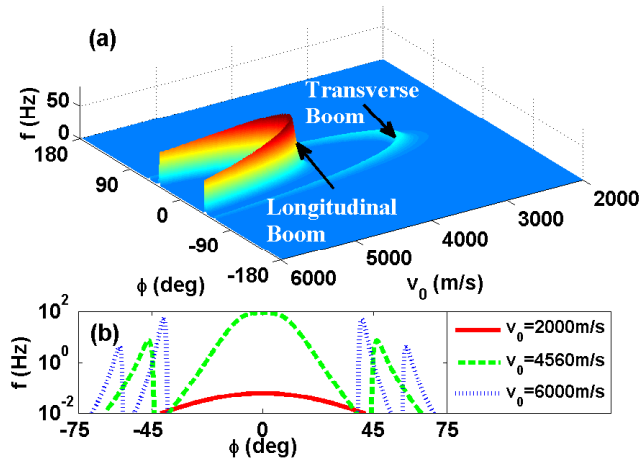} \caption{\label{fig:boom}(color online) (a) Angular distribution (azimuthal
angle $\phi$) of the kernel function $f(\theta=\frac{\pi}{2},\phi)$
for different moving velocity $v_{0}$. Here the polar angle $\theta$
is fixed at $\theta=\frac{\pi}{2}$. The parameters are $B_{0}=1$
T, $\omega_{d}=10.1$ meV, $d=20$ nm, $\phi_{v}=0$. (b) Three cross-sections
of (a) at different velocities. The red solid line has a velocity
below the speed of the transverse acoustic phonons; the green dashed
line has a velocity at the speed of longitudinal acoustic phonons,
and the blue dotted line has a velocity above the speed of longitudinal
phonons.}
\end{figure}

Quantum mechanically, we find that the moving spin qubit indeed follows
qualitatively the classical behavior. Figure \ref{fig:boom} shows
the QD speed $v_{0}$ and angle $\phi$ dependence of the kernel function
$f$ (we have chosen $\theta=\pi/2$ to maximize $f$). When $v_{0}<v_{1}$
\cite{v1}, the angular distribution is relatively flat. When taking
into account that phonon emission in spin relaxation is enabled by
spin-orbit interaction, there is a pretty strong $\sin^{3}\theta\cos^{2}\phi$
angular dependence for $f$, so that emission along directions perpendicular
to the direction of motion is suppressed. However, emissions along
all other directions are allowed. When $v_{0}\approx v_{1}$, the
angular distribution in the $xy$ plane rapidly becomes concentrated
around $\phi=0^{\circ}$, as $\phi=0^{\circ}$ is a singularity of
$w_{z}$ when $v_{0}=v_{1}$. Finally, when $v_{0}>v_{1}$, the angular
distribution in the $xy$ plane is split into two branches. Each branch
corresponds to an angle $\phi$ that gives the peak value of the kernel
function $f$. As the moving velocity gradually increases from the
subsonic regime to the supersonic regime, the kernel function gradually
concentrates into the two bifurcating angles.

The transitions through the transonic regime can be more quantitatively
seen from the cross sections given in Fig. \ref{fig:boom} (b). For
$v_{0}=2000$ m/s, which is subsonic, the kernel function is smooth
and has a small magnitude. At $v_{0}=4560$ m/s, the speed of sound
for the LA phonons, a large peak appears at $\phi=0$. Notice that
the logarithmic scale has made this peak appears to be broader than
it really is (the logrithmic scale is necessary for us to see the
subsonic value of $f$). The two side peaks are the shock waves produced
by the TA phonons, for whom the moving dot is already supersonic.
Lastly, at $v_{0}=6000$ m/s, the moving qubit is supersonic with
respect to both LA and TA phonons. Thus two sets of shock wave peaks
appear for the kernel function in this case.

\subsection{\label{sub:Cherenkov}The Cherenkov effect}

\begin{figure}
\noindent \includegraphics[width=0.95\columnwidth]{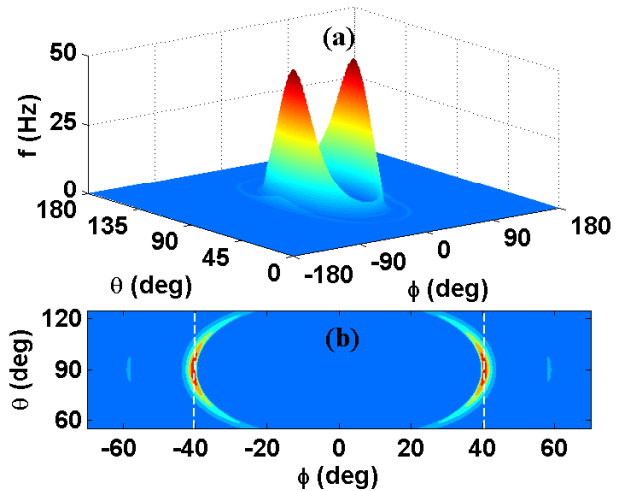} \caption{\label{fig:fj} (color online) (a) Angular distribution of the kernel
function $f(\theta,\phi)$. (b) Contour plot of $f(\theta,\phi)$
in a small region. The parameters are chosen as $v_{0}=6000$ m/s,
$B_{0}=$1 T, $\omega_{d}=0.01$ eV, $d=20$ nm, $\phi_{v}=0$.}
\end{figure}

We now examine the supersonic regime more closely, where we find clear
evidence of Cherenkov radiation of phonons from the moving spin qubit.
In Fig.~\ref{fig:fj}, we plot the kernel function $f$ as a function
of azimuthal angle $\theta$ and polar angle $\phi$ when the QD speed
is $v_{0}=6000$ m/s, larger than the speed of sound for both LA ($v_{1}$)
and TA ($v_{2}$) phonons. Clearly, \textbf{the dominant contribution
to spin relaxation is concentrated in two particular directions in
the $xy$ plane} (at $\phi\approx\pm40^{\circ}$ and $\theta\approx90^{\circ}$).
These peaks come from deformation potential interaction with LA phonons.
Two much smaller peaks appear near $\phi\approx60^{\circ}$ and $\theta\approx90^{\circ}$),
which originates from piezoelectric interaction with TA phonons.

Strong angular concentration is a typical characteristic of the Cherenkov
effect. In the current case, the emitted/absorbed phonons have a Doppler
shifted frequency of $w_{z}=\omega_{Z}/(1-\xi_{j})$. Without considering
the quantum confinement effect embodied by the cutoff functions $F_{z}$
and $F_{xy}$, the kernel function $f$ in Eq.~(\ref{eq:fj}) is
proportional to $w_{z}^{4}$ (we limit our consideration here to the
deformation potential interaction with LA phonons. The discussion
is similar when piezoelectric interaction dominates), and diverges
when $\xi_{j}=1$. Thus the angular distribution should peak along
the directions given by $\xi_{j}=1$, i.e.,
\begin{equation}
\sin\theta\cos(\phi-\phi_{v})=\frac{v_{j}}{v_{0}},\label{eq:AnglePh}
\end{equation}
which is identical to the classical Cherenkov relation in Eq.~(\ref{eq:AngleCh}).
For the parameters used to generate Fig.~\ref{fig:fj}, Eq.~(\ref{eq:AnglePh})
gives a Cherenkov angle of $\phi_{C}\approx38^{\circ}$ at $\theta=\pi/2$,
only slightly smaller (although qualitatively significant, as we discuss
in the next subsection) than the numerical value give in the figure.

In addition to the small discrepancy in the Cherenkov radiation angle,
a more significant difference between Eq.~(\ref{eq:AnglePh}) and
Fig.~\ref{fig:fj} is that the equation predicts a conical wave front
for the shock wave, while the numerical results present in the figure
have a strong two-dimensional characteristic. The bend in the peaks
in Fig.~\ref{fig:fj} can be explained by the $\sin\theta$ factor
in Eq.~(\ref{eq:AnglePh}), but the suppression of the peaks away
from $\theta=\pi/2$ cannot. We need to include all the factors in
Eq.~(\ref{eq:fj}) to explain this difference, as we will do in the
next subsection.

\subsection{Quantum correction on Cherenkov angle}

\label{sub:confinement}

While Eq.~(\ref{eq:AnglePh}) is consistent with the classical result,
our situation of the moving spin qubit is more nuanced. Indeed, $\xi_{j}=1$
would lead to a diverging phonon frequency $w_{z}$, which is clearly
unphysical. A more accurate description of the phonon Cherenkov radiation
from a moving spin qubit can only be obtained when quantum confinement
effects are taken into account.

Mathematically, quantum confinement effects are incorporated in the
cutoff functions $F_{z}$ and $F_{xy}$ in Eq.~(\ref{eq:Fc}), which
are exponentially decaying functions of the phonon frequency $w_{z}$.
Since $F_{z}$ and $F_{xy}$ decay much faster than the increase in
the power function $w_{z}^{4}$, the singularity of $w_{z}\rightarrow\infty$
is eliminated, and the peak value of the kernel function $f$ is shifted
from infinity to a large yet finite number. Physically, the cutoff
functions are simply a reflection of the phonon bottleneck effect
\cite{bottleneck1,bottleneck2}: for an electron with a finite width
$\lambda$ in its wave function, the interaction matrix element $\langle e^{iq_{\parallel}\cdot r}\rangle$
is suppressed if the phonon wave length is much smaller than $\lambda$.

For a gated QD in semiconductor nanostructures (such as a gated depletion
dot from a two-dimensional electron gas), the confinement in $z$
direction (growth direction) is much stronger than those in the $xy$
directions (in-plane directions), so that phonons are only emitted
in the $xy$ plane. This explains the more two-dimensional nature
of $f$ in Fig.~\ref{fig:fj}, instead of a conical shape. When confined
to the $xy$ plane, the kernel function $f$ is reduced to
\begin{equation}
f=\sum_{j}\frac{\hbar\omega_{Z}F_{SO}}{(m^{*}\omega_{d}^{2})^{2}}\frac{\left(2N_{w_{z}}+1\right)}{8\pi^{2}\rho_{c}v_{j}^{5}}w_{z}^{4}\cos^{2}\phi F_{xy}C_{ep},
\end{equation}
with $\xi_{j}=\frac{v_{0}}{v_{j}}\cos\phi$ (assume $\phi_{v}=0$). Now the kernel function
$f$ depends on the phonon frequency $w_{z}$ as $f\propto w_{z}^{4}\exp\left(-w_{z}^{2}\lambda^{2}/2v_{1}^{2}\right)$
(instead of $f\propto w_{z}^{4}$ when confinement effect is not included).
The peak of $f$ thus appears at
\begin{equation}
w_{z}=\frac{2v_{1}}{\lambda}.\label{eq:peakwz}
\end{equation}
Using the parameters in Fig.~\ref{fig:fj}, the peak value of $f$
occurs at $w_{z}=8.9\times10^{11}$ s$^{-1}$, while the Zeeman frequency
is $3.87\times10^{10}$ s$^{-1}$ (at $B=1$ T). With $w_{z}/\omega_{Z}\approx23$,
the phonon energy has been Doppler-shifted greatly, from about 25
$\mu$eV for $\hbar\omega_{Z}$ to nearly 600 $\mu$eV for $\hbar w_{z}$.
Recall that $w_{z}=\frac{\omega_{Z}}{1-v_{0}\cos\phi/v_{1}}$, Using
Eq.~(\ref{eq:peakwz}) it is straight forward to obtain
\begin{equation}
\phi_{C}'=\arccos\left(\frac{v_{1}}{v_{0}}-\frac{\omega_{Z}\lambda}{2v_{0}}\right).\label{eq:phicorr}
\end{equation}
Clearly, quantum confinement for the spin qubit leads to the correction
term $-\frac{\omega_{Z}\lambda}{2v_{0}}$. Using the parameters in
Fig.~\ref{fig:fj}, with an applied field of 1 T and QD speed of
$v_{0}=6000$ m/s (with the speed of sound for LA phonons at $v_{1}=4730$
m/s in GaAs), we obtain
\begin{equation}
\phi\approx\pm40^{\circ}.
\end{equation}
This is consistent with the results shown in Fig.~\ref{fig:fj},
where the peak in the kernel function $f$ appears at $\phi\approx\pm40^{\circ}$
for $\theta=\pi/2$.

In short, a combination of the Cherenkov effect and quantum confinement
leads to the results presented in Fig.~\ref{fig:fj}. The quantum
confinement for the QD that carries the spin qubit makes the phonon
emission more two-dimensional. It also suppresses the electron interaction
with higher-energy phonons, leading to a small correction to the Cherenkov
radiation angle and a significant modification to the energy of the
radiated phonons. Another consequence of the quantum confinement is
that the velocity when ``spin-relaxation boom'' occurs is also slightly
shifted, as we discuss in the next Section.

\section{Spin Relaxation in a Moving Quantum Dot}

In the last Section we have examined in detail the angular behavior
of phonon emission in the relaxation of a moving spin qubit. In this
Section we focus on the integrated effect of QD motion on spin relaxation.
We are particularly interested in how spin relaxation varies with
the speed of the QD motion and the applied magnetic field.

\begin{figure}
\noindent \begin{centering}
\includegraphics[width=1\columnwidth]{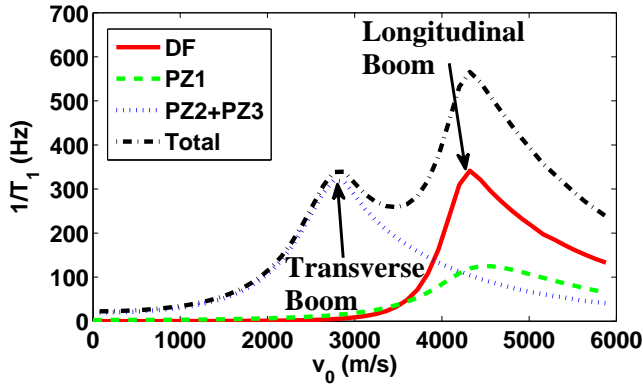}
\par\end{centering}

\caption{\label{fig:T1_boom}(color online) Spin relaxation rate $1/T_{1}$
as a function of moving velocity $v_{0}$. The red (solid), green
(dashed), blue (dotted), and black (dot-dashed) lines represent the
deformation, the longitudinal piezoelectric, the transverse piezoelectric
and the total spin relaxation rate respectively. The parameters are
chosen as $B_{0}=2$ T, $\omega_{d}=3.1$ meV, $d=20$ nm, $\phi_{v}=0$. }
\end{figure}

Classically, the drag force on an aircraft increases sharply when
the aircraft velocity approaches the sound barrier. This is the so-called
sonic boom. We find a similar behavior in the relaxation rate for
a moving spin qubit. In Fig.~\ref{fig:T1_boom}, we plot the spin
relaxation rate $1/T_{1}$ as a function of the QD speed $v_{0}$.
The curve of the total relaxation rate (black, dot-dashed) peaks at
the two sound barriers due to TA (at $v_{2}=v_{3}$) and LA phonons
(at $v_{1}$). Each peak for a single type of phonons (for example,
the red curve for the LA phonons) is similar to the Prandtl-Glauert
singularity \cite{Boom} for the classical ``sonic boom''. These
peaks can thus be named ``spin-relaxation boom''. The total relaxation
is a simple sum of contributions from LA and TA phonons. The quantum
confinement again produces some modifications to these booms. First,
the singularities are eliminated and broadened into smooth and finite
peaks. Second, the peaks are shifted downward from $v_{1}$ and $v_{2}$.

\noindent

\begin{figure}
\noindent \begin{centering}
\includegraphics[width=1\columnwidth]{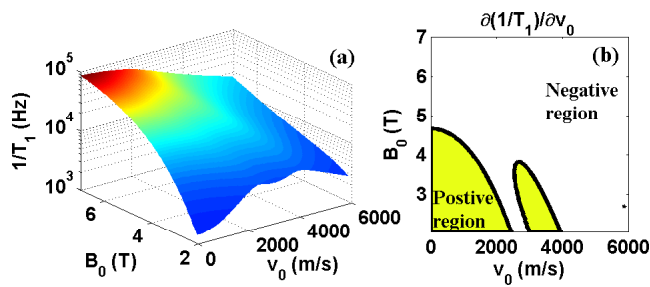}
\par\end{centering}

\caption{\label{fig:T3d_PTPV}(color online) (a) Spin relaxation rate $1/T_{1}$
as a function of magnetic field $B$ and moving speed of the quantum
dot $v_{0}$. (b) Partial derivative of $1/T_{1}$ with respect to
$v_{0}$. The white and the yellow regions indicate the partial derivative
is below and above zero respectively. The parameters are chosen as
$\omega_{d}=1.1$ meV, $d=20$ nm, $\phi_{v}=0$. }
\end{figure}

One implication of the spin relaxation booms presented above is that
a spin qubit can relax slower when the QD moves faster, as evidenced
by the curves in Fig.~\ref{fig:T1_boom}. In Fig.~\ref{fig:T3d_PTPV}
(a) we give a more comprehensive plot of this velocity dependence.
Here we plot the spin relaxation rate as a function of the moving
velocity $v_{0}$ and the magnetic field $B_{0}$. When the external
magnetic field is weak (e.g., $B_{0}\approx2$ T), the relaxation
rate increases with the moving velocity. But when the external magnetic
field is strong (e.g., $B_{0}>5$ T), the relaxation rate becomes
a decreasing function of the moving velocity. This somewhat counterintuitive
feature can be understood with the help of the ``shifted frequency''
$w_{z}$ for the emitted phonon, which depends on both the magnetic
field $B_{0}$ and the moving velocity $v_{0}$. When the moving velocity
is fast, the Doppler effect is strong, leading to the ``shifted frequency''
to get into the range where phonon bottleneck effect suppresses electron-phonon
interaction, as we have discussed in subsection \ref{sub:confinement}.
The ``shifted frequency'' $w_{z}$ also depends on the magnetic
field $B_{0}$. In a strong field, the Zeeman frequency $\omega_{Z}$
is already close to the bottleneck regime. It is thus much easier
for Doppler effect to shift the frequency higher and suppress the
phonon coupling.

\noindent

Our observation here indicates that a moving spin qubit may have an
even lower relaxation rate than a static spin qubit. While ``motional
narrowing'' is a common occurrence in spin resonance experiments
\cite{Slichter_book}, suppressing decoherence by moving a spin qubit
faster in a nanostructure setting is still an intriguing proposition.
In Fig.~\ref{fig:T3d_PTPV} (b), we plot the partial derivative of
$1/T_{1}$ with respect to $v_{0}$. Here the white region indicates
where spin relaxation can be suppressed by increasing the moving velocity,
since in this region the partial derivative of $1/T_{1}$ with respect
to $v_{0}$ is negative. On the other hand, in the yellow region the
partial derivative with respect to $v_{0}$ is positive, and relaxation
becomes faster when the spin qubit moves faster. In short, the motional
narrowing effect here shows the possibility of coherence-preserving
transportation of a spin qubit.

\section{Discussions}

Our study shows that in the supersonic regime for a moving spin qubit,
only phonons emitted or absorbed in certain directions make notable
contributions to qubit relaxation. As is shown in Fig.~\ref{fig:fj},
the kernel function $f$ for the spin relaxation rate $1/T_{1}$ is
non-zero only in certain directions in the $xy$ plane. With this
strong angular anisotropy, it is natural to consider whether we could
eliminate spin relaxation by suppressing the electron-phonon interaction
in certain directions. For example, if a phonon cavity is set up in
a certain direction, the frequencies of phonon modes in that direction
become discrete, making it possible to filter out important frequencies
and to suppress electron-phonon interaction at those frequencies.

The narrowly directional phonon emission from the moving spin qubit
may also be used as a source of phonons. Imagine an SAW provides a
stream of excited spins and they also emit phonons in one direction.
If a phonon cavity is set up along that direction and is on resonance
with the emitted phonons, it may be possible to create stimulated
emission, even lasing, of phonons in that mode \cite{Khaetskii_PRL13}.

The interesting features in phonon emission and spin relaxation we
have explored here could be useful in monitoring and detecting the
spin decoherence process. Conversely, knowing the phonon emission
angle precisely may allow continuous monitoring of the environment,
which could in turn provide more accurate information to possible
feedback operations in a quantum feedback control \cite{Wiseman_PRL_1993,Wiseman_PRA_1994}
or quantum state restoration \cite{Zhao2013,Wang2014} scheme. In
an open quantum system, the information stored in the system constantly
leaks into its environment. By measuring the environment in particular
ways, however, the lost information could be fully or partially regained.
It is thus possible to restore a system to its initial state by performing
certain operations \cite{Wang2014}. Our results about the angular
distribution of phonon emission may provide a guidance on measuring
the phonon environment: we can place the phonon detectors in selected
directions {[}precisely predicted by Eq.~(\ref{eq:phicorr}){]} since
only phonons in those directions make significant contributions to
spin relaxation.

\section{Conclusions}

In this work we have studied decoherence of a moving spin qubit caused
by phonon noise through spin-orbit interaction. We find that the QD
motion leads to Doppler shifts in the emitted/absorbed phonons by
the moving spin qubit. When the moving velocity is larger than the
sound velocity in the material, spin relaxation is dominated by phonon
emission/absorption in certain directions. The physics here is similar
to the phenomenon of classical Cherenkov radiation. We derive an explicit
formula for the quantum confinement correction to the Cherenkov angle.
We also find a ``spin-relaxation boom'' when the moving QD break
the sound barriers, in analogy to the classical sonic boom. The spin
relaxation rate peaks when the QD velocity matches the speed of sound
for a particular phonon branch. It is possible to reduce decoherence
by increasing the moving velocity. Indeed, in the supersonic regime,
the moving spin qubit may have an even lower decoherence rate than
a static qubit.

Our study on the relaxation of a moving spin qubit sheds new light
on the topic of spin-phonon interaction in a semiconductor nanostructure.
There could be significant implications in a wide range of subject
areas, from quantum coherent operations such as transferring and preserving
quantum information to more classical applications such as coherent
phonon optics.
\begin{acknowledgments}
We thank fruitful discussions with Jo-Tzu Hung, and acknowledge financial
support by US ARO (W911NF0910393) and NSF PIF (PHY-1104672).
\end{acknowledgments}
\appendix

\section{\label{sec:Heff}Derivation of the effective Hamiltonian}

An effective spin Hamiltonian, in which spin dynamics and orbital
dynamics are decoupled, can be obtained by performing a Schrieffer-Wolff
transformation to remove the SO coupling term in the full Hamiltonian
\cite{Phonon2,Phonon3,Peihao14,Golovach_PRL_2004,Huang_PRB_2013}.
Through a unitary transformation $\tilde{H}=e^{S}He^{-S}$, with $S$
given by $[H_{d}+H_{Z},S]=H_{SO}$, the SO coupling is removed to
the first order. The spin Hamiltonian is then $H_{eff}=\langle\psi|\tilde{H}|\psi\rangle$,
where $|\psi\rangle$ is the ground state of the orbital wave function.
Following the approach used in Refs.~\cite{Golovach_PRL_2004,Huang_PRB_2013},
we obtain the effective Hamiltonian in Eq.~(\ref{eq:Heff}), where
$\Omega(r,t)$ originates from the electron-phonon interaction, and
is given explicitly as
\begin{eqnarray}
\Omega_{x}(t) & = & \sum_{qj}-\frac{\beta_{-}}{\hbar\omega_{d}^{2}}\frac{iq_{y}F(q_{z})e^{-q^{2}\lambda^{2}/4}}{\sqrt{2\rho_{c}\omega_{qj}/\hbar}}\nonumber \\
 &  & \times\left(e\beta_{qj}-iq\Xi_{qj}\right)e^{iq\cdot r_{0}(t)}\left(b_{-qj}^{\text{\ensuremath{\dagger}}}+b_{qj}\right),\label{eq:Omegax}\\
\Omega_{y}(t) & = & \sum_{qj}-\frac{\beta_{+}}{\hbar\omega_{d}^{2}}\frac{iq_{x}F(q_{z})e^{-q^{2}\lambda^{2}/4}}{\sqrt{2\rho_{c}\omega_{qj}/\hbar}}\nonumber \\
 &  & \times\left(e\beta_{qj}-iq\Xi_{qj}\right)e^{iq\cdot r_{0}(t)}\left(b_{-qj}^{\text{\ensuremath{\dagger}}}+b_{qj}\right).\label{eq:Omegay}
\end{eqnarray}
The expressions here are similar to the results in Ref.~\cite{Golovach_PRL_2004},
with an additional term $e^{iq\cdot r_{0}(t)}$ due to the motion
of the quantum dot. This factor is also how Doppler effect is introduced
into the dynamics of the moving spin qubit.

\section{\label{sec:DecoRate}Derivation of the spin relaxation rate}

Given the effective Hamiltonian (\ref{eq:Heff}), the relaxation rate
can be obtained within the Bloch-Redfield theory as $\frac{1}{T_{1}}=J_{XX}^{+}(\omega_{Z})+J_{YY}^{+}(\omega_{Z})$
\cite{Phonon3}, where $\omega_{Z}=g\mu_{B}/\hbar$ is the Zeeman
frequency. The tensors $J_{XX}^{+}$ and $J_{YY}^{+}$ are correlations
of the effective magnetic noise (from the phonons through the SO interaction),
\begin{equation}
J_{ij}^{+}(\omega)=\frac{g^{2}\mu_{B}^{2}}{4\hbar^{2}}{\rm Re}\int_{-\infty}^{\infty}\left\langle \left\{ \delta B_{i}(0),\delta B_{j}(t)\right\} \right\rangle e^{-iwt}dt\,.\label{eq:Jij}
\end{equation}
These correlation functions are expressed in a rotated $XYZ$ coordinate
system, where the $Z$ axis is along the direction of the applied
field $B_{0}$. Due to this rotation, a magnetic-angular dependence
term $F_{SO}$ appears in the expression of the relaxation rate \cite{Phonon3,Golovach_PRL_2004}.
With the magnetic noise from phonons, the relaxation rate takes the
form
\begin{eqnarray}
\frac{1}{T_{1}} & = & F_{SO}\omega_{Z}^{2}{\rm Re}\int_{-\infty}^{\infty}dt\sum_{j}\int d\theta\int d\phi\int d\omega_{j}e^{-i\omega_{Z}t}\frac{\omega_{j}^{3}}{v_{j}^{5}}\nonumber \\
 &  & \times\frac{\sin^{3}\theta\cos^{2}\phi}{(\hbar\omega_{d}^{2})^{2}}\frac{|F(\frac{\omega_{j}}{v_{j}}\cos\theta)|^{2}e^{-q^{2}\lambda^{2}/2}}{(2\pi)^{3}2\rho_{c}/\hbar}C_{ep}A,\label{eq:relaxation rate}
\end{eqnarray}
where
\begin{align}
A\equiv & \left\langle b_{qj}^{\dagger}b_{qj}e^{-i\omega_{qj}(1-\xi_{j})t}+b_{qj}b_{qj}^{\text{\ensuremath{\dagger}}}e^{i\omega_{qj}(1-\xi_{j})t}\right\rangle \nonumber \\
 & +\left\langle b_{qj}^{\dagger}b_{qj}e^{i\omega_{qj}(1-\xi_{j})t}+b_{qj}b_{qj}^{\text{\ensuremath{\dagger}}}e^{-i\omega_{qj}(1-\xi_{j})t}\right\rangle .
\end{align}
Compared with Ref.~\cite{Golovach_PRL_2004}, the phonon frequency
here is shifted by the factor $(1-\xi_{j})$, which is a result of
the Doppler effect. After performing the time integral in Eq.~(\ref{eq:relaxation rate})
we obtain the relaxation rate (\ref{eq:T1}) and the kernel function
(\ref{eq:fj}). When the moving velocity approaches zero, the relaxation
rate here reduces exactly to the result shown in Ref.~\cite{Huang_PRB_2013}.


\begin{thebibliography}{10}
\bibitem{v1} For the parameters used in this paper, deformation potential
coupling to LA phonons provides the dominant spin relaxation channel.
Therefore the critical velocity for Cherenkov effect is approximately
$v_{1}$. For a case when piezoelectric potential is dominant or when
both interactions are important, there would be two critical velocities,
determined by the two types of phonons.
\end{thebibliography}
\end{document}